\documentclass[prl,aps,groupedaddress,showpacs,twocolumn]{revtex4-1}
\usepackage{bm}% bold math
\usepackage{graphicx}

\begin{document}
\title{Two types of Dirac-cone surface states on (111) surface of topological crystalline insulator SnTe}
\author{Y. Tanaka,$^1$ T. Shoman,$^1$ K. Nakayama,$^1$ S. Souma,$^2$ T. Sato,$^1$ T. Takahashi,$^{1,2}$ M. Novak,$^3$ Kouji Segawa,$^3$ and Yoichi Ando$^3$}
\affiliation{$^1$Department of Physics, Tohoku University, Sendai 980-8578, Japan\\
$^2$WPI Research Center, Advanced Institute for Materials Research,
Tohoku University, Sendai 980-8577, Japan\\
$^3$Institute of Scientific and Industrial Research, Osaka University, Ibaraki, Osaka 567-0047, Japan}

\date{\today}

\begin{abstract}
We have performed angle-resolved photoemission spectroscopy (ARPES) on the (111) surface of the topological crystalline insulator SnTe. Distinct from a pair of Dirac-cone surface states across the $\bar{X}$ point of the surface Brillouin zone on the (001) surface, we revealed two types of Dirac-cone surface states each centered at the $\bar{\Gamma}$ and $\bar{M}$ points, which originate from the bulk-band inversion at the L points. We also found that the energy location of the Dirac point and the Dirac velocity are different from each other. This ARPES experiment demonstrates the surface states on different crystal faces of a topological material, and it elucidates how mirror-symmetry-protected Dirac cones of a topological crystalline insulator show up on surfaces with different symmetries.
\end{abstract}

\maketitle
\section{INTRODUCTION}
 Topological insulators (TIs) embody a topologically nontrivial quantum state of matter where an insulating bulk with an inverted energy gap induced by a strong spin-orbit coupling is necessarily accompanied by gapless surface states (SSs) that are protected by time-reversal symmetry \cite{Hasan:2010ku, Qi:2011hb, Ando:2013fg}. The discovery of TIs triggered the search for new types of topological materials protected by other symmetries \cite{Schnyder:2008ez, Kitaev:2009vc, Ran:2010wx, Mong:2010wd, Li:2010eo}, and such investigations have recently been extended to topological crystalline insulators (TCIs) \cite{Fu:2011ia, Hsieh:2012cn} in which gapless SSs are protected by a point-group symmetry of the crystalline lattice, in particular, mirror symmetry (reflection symmetry) of the crystal. The narrow-gap IV-VI semiconductor SnTe with a rock-salt structure has been theoretically predicted to be a first example of TCI \cite{Hsieh:2012cn}, where the band inversion takes place at the L points of the fcc bulk Brillouin zone (BZ) [see, \textit{e.g.}, Fig. 1(a)], corresponding to an even number (four) of inversion at time-reversal invariant momenta (TRIMs). This leads to the trivial $\mathcal{Z}_2$ topological invariant (0;000), but its mirror symmetry gives rise to a nontrivial mirror Chern number $n_{\mathcal{M}} = -2$. The TCI phase has been experimentally verified by angle-resolved photoemission spectroscopy (ARPES) on the (001) surface of SnTe \cite{Tanaka:2012kf}, Pb$_{1-x}$Sn$_x$Te ($x < 0.25$) \cite{Xu:2012bm,Tanaka:2013kl} and Pb$_{0.73}$Sn$_{0.27}$Se \cite{Dziawa:2012hx, Wojek:2013bh}, in which the Dirac-cone SSs are located at momenta slightly away from the TRIM $\bar{X}$ point but along the (110) mirror plane of the crystal, to produce a characteristic ``double Dirac-cone'' structure. While a non-zero mirror Chern number theoretically guarantees the existence of topologically protected gapless SSs on \textit{any surface containing a mirror plane} \cite{Liu:2013wu, Safaei:2013js}, it has not been experimentally verified whether the topologically protected Dirac-cone SSs indeed exist on other surface-plane orientations like (111) and how the surface band structure depends on the symmetry of each surface orientation. This point is of crucial importance not only for understanding the interplay between the crystal symmetry and the topological protection in TCIs, but also for clarifying a possible surface-orientation dependence of the topological SSs in topological materials in general. It is worth mentioning that, because of the experimental difficulty in preparing clean and flat surfaces with different orientations, it has not been possible to experimentally compare topological SSs on different surfaces of a topological material \cite{note1}.
 
 In this article, we report our ARPES study of SnTe single crystals which turned out to be cleavable along both the (001) and (111) crystal planes \cite{Sato:2013br}. Our synchrotron-based ARPES experiments on the (111) surface with variable photon energies unambiguously demonstrate that there exist two types of Dirac-cone SSs centered at the $\bar{\Gamma}$ and $\bar{M}$ points, whose characteristics are different from the Dirac-cone SSs on the (001) surface in some important aspects. We discuss the implications of our experimental findings in relation to the band-structure calculations \cite{Liu:2013wu,Safaei:2013js}.
  
\section{EXPERIMENT}
 High-quality single crystals of SnTe used in this study were grown by a vapor-transport method \cite{Sato:2013br} using high-purity elements of Sn (99.99\%) and Te (99.999\%). All those vapor-grown crystals are naturally faceted and show good crystallinity in x-ray Laue analyses. ARPES measurements were performed with the MBS-A1 and VG-Scienta SES2002 electron analyzers with tunable synchrotron lights at the beamline BL-7U at UVSOR as well as at the beamline BL28A at Photon Factory (KEK). To excite photoelectrons, we used the linearly polarized lights of 12-40 eV and the circularly polarized lights of 50-100 eV, at UVSOR and Photon Factory, respectively. The energy and angular resolutions were set at 10-30 meV and 0.2$^\circ$, respectively. Samples having reasonably large (111) facets were cleaved \textit{in-situ} in an ultrahigh vacuum of $1\times10^{-10}$ Torr along the (111) crystal plane by knocking a post glued to the (111) face, and the sample temperature was always kept at $T = 30$ K. A shiny mirror-like surface was obtained after cleaving the samples as in the case of the (001) surface \cite{Tanaka:2012kf}, and the surface orientation was confirmed by the x-ray Laue diffraction which clearly displayed the six-fold symmetric pattern. All the ARPES data were recorded within 7 hours after cleaving, during which we did not observe any signatures of surface degradation. As seen in Fig. 1 (a), the (110) mirror plane of the bulk BZ lies along the $\bar{\Gamma}\bar{M}$ line in the (111) surface BZ. One of the bulk L points at which the band inversion occurs is projected onto the surface $\bar{\Gamma}$ point, while other three L points are projected onto the surface $\bar{M}$ points.
 
 \begin{figure}
 \includegraphics[width=3.4in]{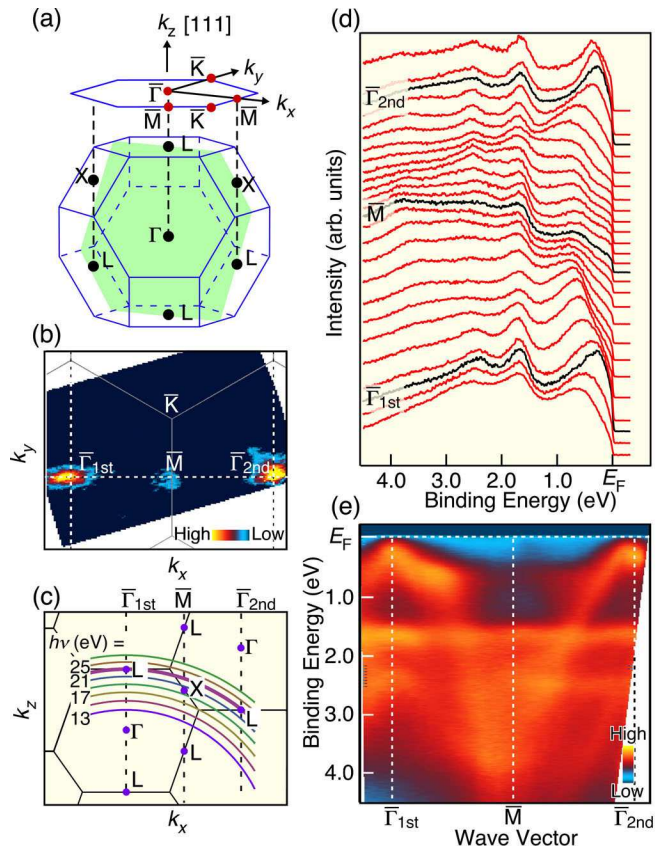}
\caption{(Color online) Bulk fcc Brillouin zone (BZ) and corresponding hexagonal (111) surface BZ of SnTe. Green shaded area represents the (110) mirror plane of the rock-salt structure. (b) ARPES intensity mapping at $E_{\rm F}$ at $T = 30$ K on the (111) surface plotted as a function of in-plane wave vector; this intensity is obtained by integrating the spectra within $\pm$10 meV of $E_{\rm F}$. (c) Bulk BZ in the $k_x$-$k_z$ plane, together with measured cuts for $h\nu = 13$-27 eV. The inner potential value is estimated to be 6.9 eV from the normal-emission ARPES measurements. (d) EDCs in the VB region along the $\bar{\Gamma}\bar{M}$ line at $h\nu = 23$ eV. (e) ARPES intensity in the VB region along the $\bar{\Gamma}\bar{M}$ line plotted as a function of wave vector and binding energy.}
\end{figure}

\section{RESULTS AND DISCUSSION}
 Figure 1(b) displays the ARPES intensity at $E_{\rm F}$ on the (111) surface of SnTe plotted as a function of in-plane wave vector measured with the photon energy of $h\nu = 23$ eV; this measurement was performed along the cut indicated by a thick curve in the $k_x$-$k_z$ plane in Fig. 1(c) which nearly passes the L points of the first and second bulk BZ as well as the X point. One can immediately recognize from Fig. 1(b) the bright intensity centered at the first and second $\bar{\Gamma}$ points as well as the relatively weak intensity at the $\bar{M}$ point, which reflects the periodicity of the (111) surface. The observed intensity at $E_{\rm F}$ originates from the holelike band near $E_{\rm F}$ seen in the valence-band (VB) energy distribution curves [EDCs; Fig. 1(d)] and the ARPES intensity plot [Fig. 1(e)], and it originates from a combination of the SS and the highest occupied bulk VB consisting mainly of the Te 5$p$ orbitals \cite{Liu:2013wu,Safaei:2013js,Tung:1969bn,Melvin:2001fb,Littlewood:2010ec}.
 
  \begin{figure*}
 \includegraphics[width=6.4in]{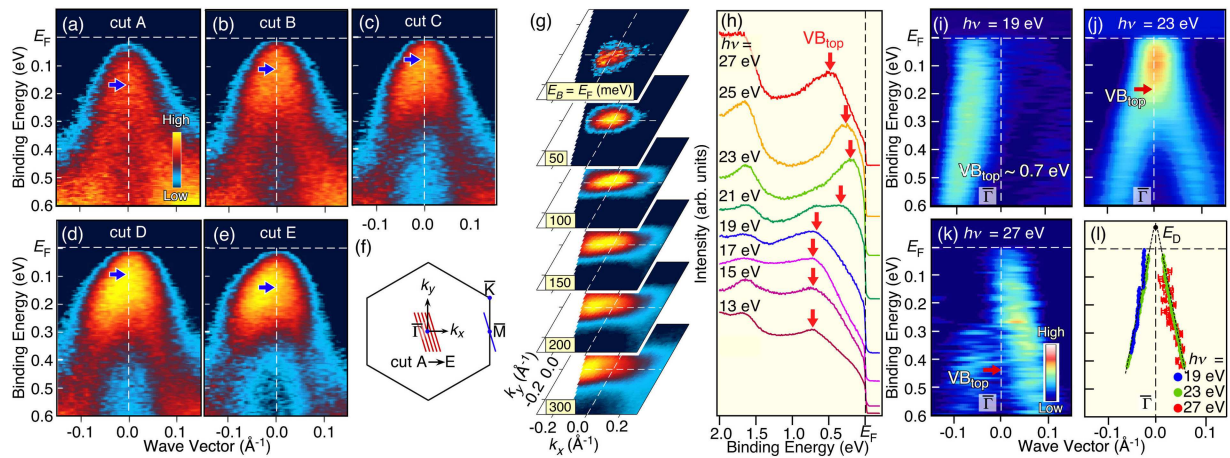}
\caption{(Color online) (a)-(e) Near-$E_{\rm F}$ ARPES intensity of SnTe plotted as a function of the wave vector and binding energy at $h\nu =$ 23 eV measured along several cuts (A-E) around the $\bar{\Gamma}$ point shown by red lines in the surface BZ in (f). Blue arrow indicates the top of the band. (g) ARPES intensity maps around the $\bar{\Gamma}$ point as a function of two-dimensional wave vector at various binding energies, measured at $h\nu = $ 23 eV. (h) Photon-energy dependence of normal-emission EDCs. Red arrow indicates the peak position. (i)-(k) Intensity plots of the second derivatives of the momentum distribution curves (MDCs) across the cut crossing the $\bar{\Gamma}$ point for $h\nu = 19$, 23 and 27 eV. Red arrow denotes the top of the bulk VB which is determined in (h). (l) Comparison of the band dispersions for different photon energies extracted by fitting the MDCs for (i)-(k) with two Lorentzians \cite{SM}; error bars reflect the uncertainties originating from the momentum resolution and the standard deviation in the peak positions of MDCs. Dotted lines represent the linear extrapolations of the band. $E_{\rm D}$ denotes the energy position of the Dirac point.}
\end{figure*}

 To elucidate the near-$E_{\rm F}$ band dispersion in detail, we have performed ARPES measurements for several cuts surrounding the $\bar{\Gamma}$ point of the first surface BZ, and the result is shown in Figs. 2(a)-(e). Along cut A [for $\bf{k}$ location, see Fig. 2(f)], the observed band has its top at the binding energy of 0.17 eV. On moving from cut A to C, the band (marked by blue arrow) gradually approaches $E_{\rm F}$ (cuts A-C) and then disperses back towards higher binding energy (cuts D-E). This establishes the cone-shaped dispersion in two-dimensional (2D) $\bf{k}$ space, as also supported by the ARPES intensity maps as a function of 2D wave vector for various binding energies in Fig. 2(g) which signify a conical intensity distribution (note that the intensity appears to be slightly elongated along the $k_x$ axis likely due to the matrix-element effect of photoelectron intensity). To distinguish the bulk and surface bands, we have performed photon-energy-dependent ARPES measurements, and the corresponding normal-emission EDCs are displayed in Fig. 2(h). One can immediately find a peak displaying a sizable dispersion upon $h\nu$ variation (marked by red arrow), which is attributed to the top of the bulk VB. Besides this prominent bulk band, we find in the spectra near $E_{\rm F}$ a weaker feature that disperses linearly with momentum; this feature is best seen in the plots of the second derivative of the momentum distribution curves (MDCs), as shown in Figs. 2(i)-(k) for $h\nu = 19, 23$, and 27 eV, respectively, along a cut crossing the $\bar{\Gamma}$ point (the raw MDC curves are shown in the supplemental material \cite{SM}). Clearly, this linear dispersion approaches and touches $E_{\rm F}$ as if a Dirac point is located above $E_{\rm F}$. Note that the ARPES intensity is strongly modulated by the matrix-element effect as in the case of the (001) surface \cite{Tanaka:2012kf} which makes the second-derivative peaks to be photon-energy dependent. Nevertheless, the near-$E_{\rm F}$ band dispersion [Fig. 2(l)] extracted from the numerical fittings of the MDCs with two Lorentzians \cite{SM} reasonably overlaps with each other within the experimental uncertainties irrespective of the energy location of the bulk VB top (that ranges from 0.2 to 0.7 eV), which indicates its surface origin. This result strongly suggests the existence of a Dirac-cone SS centered at the $\bar{\Gamma}$ point, whose energy position of the Dirac point (Dirac energy; $E_{\rm D}$) lies above $E_{\rm F}$.
  
 \begin{figure}[h]
 \includegraphics[width=3.3in]{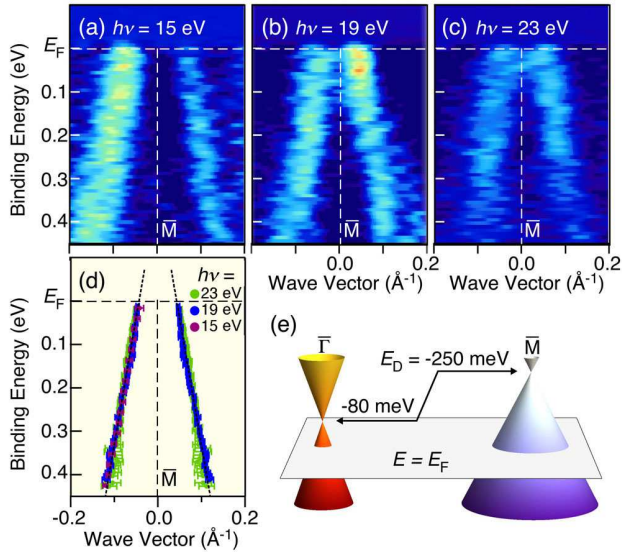}
 \caption{(Color online) (a)-(c) Intensity plots of the second derivatives of the MDCs across the cut crossing the $\bar{M}$ point [blue line in Fig. 2(f)] for $h\nu = 15$, 19 and 23 eV. (d) Comparison of the band dispersions for different photon energies extracted by tracking the peak position of MDCs by numerical fittings \cite{SM}. (e) Schematic picture of the Dirac cones located at the $\bar{\Gamma}$ and $\bar{M}$ points. The estimated difference of the $E_{\rm D}$ value between the $\bar{\Gamma}$ and $\bar{M}$ points is $\sim$170 meV.}
\end{figure}
 
Next we focus on the electronic states around the $\bar{M}$ point. Figures 3(a)-(c) show ARPES-derived band dispersion along the cut crossing the $\bar{M}$ point [blue line in Fig. 2(f)], for three representative photon energies of $h\nu = 15, 19$ and 23 eV. We identify a linearly dispersive holelike band crossing $E_{\rm F}$ in Figs. 3(a)-(c), in which the slight $h\nu$ dependence in the dispersions is an artifact of the second-derivative method applied to the raw spectra. In fact, as shown in Fig. 3(d), the true band dispersion extracted from the numerical fittings to the MDCs overlaps well with each other in the near-$E_{\rm F}$ region, indicating its 2D nature. Combining this observation with the fact that the measured $k_z$ values were selected to be apart from the bulk L point [see Fig. 1(c)], one may conclude that the observed linearly dispersive band originates from a Dirac-cone SS centered at $\bar{M}$, similarly to the case of the $\bar{\Gamma}$-centered Dirac cone. This indicates that there exist four Dirac cones in the surface BZ, one centered at $\bar{\Gamma}$ and three at $\bar{M}$, consistent with the trivial nature of this material in terms of the $\mathcal{Z}_2$ topology. We note, however, that the ARPES intensity of the $\bar{M}$-centered Dirac cone is much weaker than that at $\bar{\Gamma}$ [as one can see in Fig. 1(b)], which made it technically difficult to elucidate the anisotropy of the $\bar{M}$-centered Dirac cone in the present experiment.

We have estimated the $E_{\rm D}$ value from the linear extrapolation of the band dispersion in Figs. 2(l) and 3(d) to be 80 and 250 meV above $E_{\rm F}$ for the $\bar{\Gamma}$- and $\bar{M}$-centered Dirac cones, respectively, as shown in Fig. 3(e), implying that the fillings of the hole-doped Dirac cones are different from each other. It is noted that the band-structure calculations on the (111) surface of SnTe \cite{Liu:2013wu} and Pb$_{0.4}$Sn$_{0.6}$Te \cite{Safaei:2013js} also predict two different Dirac-cone SSs centered at the $\bar{\Gamma}$ and $\bar{M}$ points (which have opposite mirror eigenvalues), consistent with the present ARPES result. Our data resemble the calculated SSs for the Sn-terminated surface whose Dirac points are situated close to the VB edge, but are obviously different from the Te-terminated counterpart exhibiting no Dirac-cone dispersion near the VB \cite{Liu:2013wu,Safaei:2013js}. This implies that the ARPES measurement preferentially observes the Sn-terminated surface, for which a downward band bending is expected because of the partially polar nature of the (111) surface of this material \cite{Taskin:2013vl}. Although the cleaved surface should also contain Te-terminated terraces, it is expected that the Te-terminated surface has an upward band bending, which causes its surface state to be inaccessible. Note that the Sn-Te bonding is partially ionic, and thus the Sn (Te)-terminated surface is expected to be positively (negatively) charged.

 Figure 4 compares the observed Dirac-cone SSs between the (001) \cite{Tanaka:2012kf} and (111) surfaces. The SSs for the (001) surface consists of a double-Dirac-cone structure whose Dirac points are located away from the TRIM $\bar{X}$ point, but on the projection of the mirror plane, \textit{i.e.}, the mirror-symmetric line (green line) \cite{Tanaka:2013kl}. This double-Dirac-cone structure arises from the hybridization of two Dirac cones, because two L points (which are each responsible for a surface Dirac cone due to the band inversion at the L point) are projected onto the same $\bar{X}$ point. On the other hand, for the (111) surface, always a single L point is projected onto either the $\bar{\Gamma}$ or $\bar{M}$ points to produce a single Dirac cone centered at those TRIMs, and no such hybridization takes place. As a result, the Dirac point (which is on the mirror-symmetric line) coincides with the TRIM $\bar{\Gamma}$ or $\bar{M}$.
  
 \begin{figure}[b]
 \includegraphics[width=3.35in]{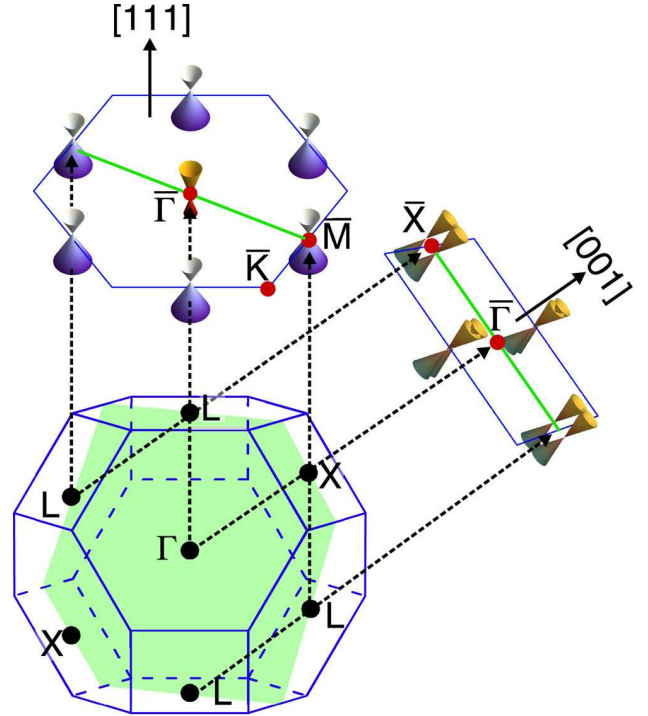}
 \caption{(Color online) Schematic picture of the Dirac-cone SS on two different surface planes of (001) \cite{Tanaka:2012kf,Tanaka:2013kl} and (111). Green shaded area and green line represent the (110) mirror plane and the mirror-symmetric line on the surface BZ, respectively.}
 \end{figure}

 Now we discuss quantitative differences between the two Dirac cones on the (111) surface. The velocity of the band at the Dirac point (the Dirac velocity, $v_{\rm D}$), estimated from the linear extrapolation of the band dispersion in Figs. 2(l) and 3(d), is 8.4$\pm$0.6 and 5.8$\pm$0.4 eV\AA\ for the $\bar{\Gamma}$- and $\bar{M}$-centered Dirac cones, respectively [note that it is difficult to estimate the momentum anisotropy of the $v_{\rm D}$ value at each Dirac cone since the $k$ resolution is insufficient along the cut perpendicular to the cuts in Fig. 2(f)]. To the best of our knowledge, the $v_{\rm D}$ value of 8.4 eV\AA\ ($1.3\times10^8$ cm/s) is the highest among any known TIs and TCIs, and is comparable to graphene \cite{Novoselov:2005kj,Bostwick:2006hn}, which would be useful for utilizing high-mobility Dirac carriers in a gated device. The relatively large $v_{\rm D}$ value of the $\bar{\Gamma}$-centered Dirac cone is related to the orientation of the constant energy ellipsoids of the bulk VB located at the L points \cite{Liu:2013wu,Safaei:2013js}; namely, the direction of the major axis of the ellipsoid whose center is projected to the $\bar{\Gamma}$ point is perpendicular to the (111) surface plane, whereas the axis is tilted with respect to the (111) plane in the case of the ellipsoids whose centers are projected to the $\bar{M}$ points. This leads to the narrower (wider) $k$-space distribution of the bulk-band projection onto the (111) surface around the $\bar{\Gamma}$ ($\bar{M}$) point. Since the SSs are known to evolve smoothly from the edge of the bulk-band projection, it is naturally expected that the $\bar{\Gamma}$-centered Dirac cone has a larger band velocity, as can be seen in the band-structure calculations \cite{Liu:2013wu,Safaei:2013js}. It should also be noted here that a finite difference in the $E_{\rm D}$ values between the two Dirac cones [Fig. 3(e)] has been predicted in the band calculations \cite{Liu:2013wu,Safaei:2013js}, although the relative $E_{\rm D}$ location of the $\bar{\Gamma}$- and $\bar{M}$-centered Dirac cones is reversed between the calculations and the present ARPES experiment. While the origin of this difference is unclear at present, it may be related to the limits of the precise quantitative predictions of the Dirac-cone surface states based on $ab$ $initio$ band calculations.

\section{SUMMARY}
   
 In summary, we have reported high-resolution ARPES results on the (111) surface of SnTe, and clarified the existence of two types of Dirac-cone SSs centered at the  $\bar{\Gamma}$ and $\bar{M}$ points, which originate from the bulk-band inversion at the L points. We also found that the characteristics of the Dirac cones, such as the Dirac energy and the Dirac velocity, are markedly different between the two Dirac cones. The present ARPES result on the (111) surface, together with our previous study on the (001) surface, allows us to directly compare topological surface states on different crystal faces of a topological material and establishes an essential role of the crystal mirror symmetry and the bulk-band inversion in realization of the TCI phase.

\begin{acknowledgements}
We thank L. Fu and A. Taskin for useful discussions. We also thank M. Nomura,  K. Honma, K. Ono, H. Kumigashira, M. Matsunami, and S. Kimura for their help in the ARPES measurements. This work was supported by JSPS (KAKENHI 23224010, 24654096, 25287079, and 25220708), MEXT of Japan (Innovative Area ``Topological Quantum Phenomena''), AFOSR (AOARD 124038), the Mitsubishi Foundation, KEK-PF  (Proposal No. 2012S2-001), and UVSOR (Proposal No. 24-536).  \end{acknowledgements}

\newpage
\onecolumngrid
\begin{center}
{\large Supplemental Materials for Two types of Dirac-cone surface states on (111) surface of\\
topological crystalline insulator SnTe}

Y. Tanaka,$^1$ T. Shoman,$^1$ K. Nakayama,$^1$ S. Souma,$^2$ T. Sato,$^1$, T. Takahashi,
\newline
$^{1,2}$ M. Novak,$^3$ Kouji Segawa,$^3$ and Yoichi Ando$^3$

{\footnotesize
$^1${\it Department of Physics, Tohoku University, Sendai 980-8578, Japan}
\newline
$21${\it WPI Research Center, Advanced Institute for Materials Research,
Tohoku University, Sendai 980-8577, Japan}
\newline
$^3${\it Institute of Scientific and Industrial Research, Osaka University, Ibaraki, Osaka 567-0047, Japan}
}

\end{center}

\renewcommand{\thefigure}{S\arabic{figure}}
\setcounter{figure}{0}

\subsection{S1. Raw ARPES data for the $\bar{\Gamma}$- and $\bar{M}$-centered Dirac cones}

Figures S1(a) and S1(b) display momentum distribution curves (MDCs) of SnTe on the (111) surface plotted for various binding energies ($E_{\rm B}$), which were used for obtaining second-derivative plots of MDCs for the $\bar{\Gamma}$-centered Dirac cone [Figs. 2(i)-(k) of the text] and the $\bar{M}$-centered Dirac cone [Figs. 3(a)-(c) of the text], respectively. For all the photon energies, we identify a linearly dispersive band crossing the Fermi level ($E_{\rm F}$), in which the relative intensity of the left- and right-hand-side features appears to be strongly dependent on the photon energy due to the matrix-element effect of the photoelectron intensity. We have estimated the peak position of the band by numerical fittings with two Lorentzians, and confirmed that the band dispersion is insensitive to the $h\nu$ variation within the experimental uncertainties, suggesting its surface origin.

 \begin{figure}[h]
 \includegraphics[width=3.4 in]{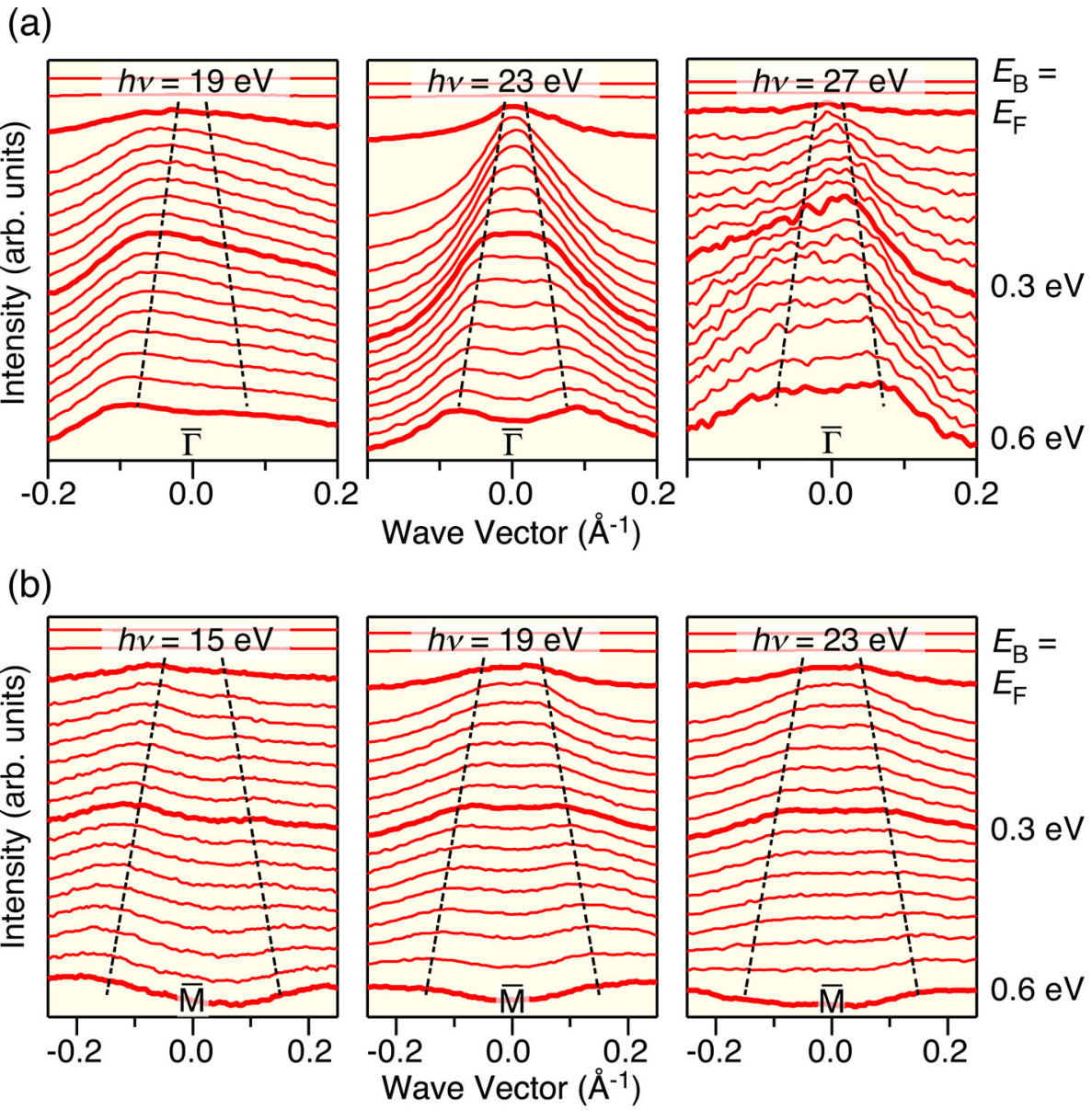}
\caption{(a), (b) Binding-energy dependence of the MDCs for the $\bar{\Gamma}$- and $\bar{M}$-centered Dirac cones, respectively, plotted for representative photon energies. Dotted lines represent the linear fittings to the bands obtained in Figs. 2(l) and 3(d) of the main text.}
\end{figure}

%\renewcommand{\thefigure}{%
   % \thesection.\arabic{figure}}
    %\@addtoreset{figure}{section}
    
%\bigskip

%\newpage

%\end{widetext}

\end{document}